\documentclass[aps,reprint,twocolumn,pra,superscriptaddress,floatfix,nofootinbib]{revtex4-1}
\usepackage{amssymb,amsmath,amsfonts} \usepackage{epsfig,graphicx}
\usepackage[colorlinks=true,linkcolor=blue,citecolor=blue]{hyperref}

\usepackage{soul}
\usepackage{color}
\setulcolor{red}
\setstcolor{red}

\newcommand\braket[1]{\left\langle\textstyle{#1}\right\rangle}

\begin{document}

\title{Fokker-Planck formalism approach to Kibble-Zurek scaling laws and non-equilibrium dynamics}
\author{Ricardo Puebla}\email{ricardo.puebla@uni-ulm.de}
\affiliation{Institut f\"{u}r Theoretische Physik and IQST,
  Albert-Einstein Allee 11, Universit\"{a}t Ulm, 89069 Ulm, Germany}
\author{Ramil Nigmatullin}\email{ramil.nigmatullin@sydney.edu.au}
\affiliation{Complex Systems Research Group, Faculty of Engineering and IT, The University of Sydney, Sydney, NSW 2006, Australia}
\author{Tanja E. Mehlst\"aubler}\email{tanja.mehlstaeubler@ptb.de}
\affiliation{Physikalisch-Technische Bundesanstalt, Bundesallee 100, 38116 Braunschweig, Germany}
\author{Martin B. Plenio}\email{martin.plenio@uni-ulm.de} \affiliation{Institut f\"{u}r Theoretische Physik and IQST, Albert-Einstein Allee 11, Universit\"{a}t Ulm, 89069 Ulm, Germany}
\date{\today}
\begin{abstract}
We study the non-equilibrium dynamics of second-order phase transitions
in a simplified Ginzburg-Landau model using the Fokker-Planck
formalism. In particular, we focus on deriving the Kibble-Zurek
scaling laws that dictate the dependence of spatial correlations on
the quench rate. In the limiting cases of overdamped and underdamped
dynamics, the Fokker-Planck method confirms the theoretical
predictions of the Kibble-Zurek scaling theory.  The developed
framework is computationally efficient, enables the prediction of
finite-size scaling functions and is applicable to microscopic models as
well as their hydrodynamic approximations. We demonstrate this
extended range of applicability by analyzing the non-equilibrium
linear to zigzag structural phase transition in ion Coulomb crystals
confined in a trap with periodic boundary conditions.
\end{abstract}

\maketitle

\section{Introduction}
Non-equilibrium dynamics involving critical phenomena, such as phase
transitions, is an important area of statistical physics~\cite{Landau80}.
The physical phenomena that arise when traversing a symmetry breaking
second-order phase transitions at finite rate are of particular interest.
Specifically, symmetry breaking at finite rate promotes the formation of
non-equilibrium excitations that can stabilize forming topological defects
in a process known as the Kibble-Zurek (KZ) mechanism~\cite{Kibble:76}.
When the quench is performed at finite rates the symmetry is broken locally,
and spatially separated regions can select different symmetry-broken
states within the ground state manifold, which results in defects forming
at spatial locations where phases of different symmetries meet. A major
achievement of KZ theory is the prediction that the average number of defects
exhibits a power-law dependence on the quench rate, whose scaling exponents
are determined by the equilibrium critical exponents of the phase transition.

KZ mechanism has been studied in a number of experiments (see~\cite{delCampo:14}
for a recent review). While the standard KZ argument applies in spatially
homogeneous systems, in some experimental systems, such as Bose-Einstein
condensates~\cite{Sadler:06,Weiler:08,Lamporesi:13} and ion Coulomb crystal
\cite{Pyka:13,Ulm:13} inhomogeneities, as well as finite-size
effects need to be accounted for. Thus measured scaling may not agree
with the prediction of KZ scaling exponents in the thermodynamic
limit. In such cases numerical simulations are particularly valuable
tools for gaining insights into the non-equilibrium dynamics. Simulations
of KZ experiments typically involve the numerical evaluation of many stochastic
trajectories to allow for the calculation of an accurate
estimate of any statistical quantity, including the expected density of
defects. The tracking of stochastic trajectories of individual quench
realizations, followed by averaging over the obtained ensemble is
known as the \emph{Langevin} approach of stochastic thermodynamics. In
stochastic thermodynamics, there exists the different but equivalent
approach of studying expectations of observables known as the
\emph{Fokker-Planck} approach~\cite{Risken:84}. Fokker-Planck
equations are deterministic partial differential equations specifying
the time-evolution of the probability distribution of the
configuration of the system that interacts with a Markovian heat
bath. Thus the Fokker-Planck approach aims to solve the non-equilibrium
dynamics problem at the ensemble rather than at the individual
realization level, as is the case in the Langevin approach. The aim of
the current paper is to apply the Fokker-Planck formalism to the KZ
problem. We develop the Fokker-Planck method for the KZ problem and
show that it can reproduce the known non-equilibrium scaling laws. The
advantages of this approach include a computationally fast evaluation
of the scaling laws and access to numerically exact probability
distributions.

The paper is structured as follows. In Section~\ref{sec:noneq}, we
formulate the problem by introducing the Ginzburg-Landau model of
phase transition, the equations of motion within the Langevin and
Fokker-Planck formulation and the observables relevant in the context
of KZ scenario. In Section~\ref{sec:smo} and~\ref{sec:kram}, we solve
the Fokker-Planck equations in, respectively, the overdamped regime
and underdamped regime. In Section~\ref{sec:crys}, we apply the method
to a non-equilibrium structural phase transition between linear and
zigzag configurations in Coulomb crystals.

\section{Non-equilibrium dynamics in Ginzburg-Landau theory}
\label{sec:noneq}

Ginzburg-Landau (GL) theory provides a good model of second-order
phase transitions. We consider a scalar one-dimensional order
parameter $\phi(x,t)$. In the GL theory of second-order phase
transitions, the free energy of the systems is given by
\begin{equation}
\label{eq:freener}
\mathcal{F}=\frac{1}{2}\int dx \left[h^2(\partial_x\phi)^2
  +V(\phi)\right],
\end{equation}
where the Ginzburg-Landau potential $V(\phi)$ reads
\begin{equation}
\label{eq:GLpot}
V(\phi)=\frac{\varepsilon}{2} \phi^2+\frac{g}{4}\phi^4.
\end{equation}
The constants $g$ and $h$ are parameters of the model that depend on
the microscopic structure of the system. The parameter $\varepsilon$
quantifies the distance to the critical point of the phase transition,
located at $\varepsilon=\varepsilon_c=0$. The model specified by
Eqs.~(\ref{eq:freener}) and~(\ref{eq:GLpot}) is ubiquitous in physics
as it describes a symmetry-breaking phase transition: the position
of the minimum of $V(\phi)$ changes from being found at $\phi=0$
for $\varepsilon>0$ to two energetically equivalent choices at $\phi=\pm
\sqrt{-\varepsilon/g}$ for $\varepsilon<0$.

The main purpose of the present article resides in analyzing the
non-equilibrium dynamics resulting from the finite-rate symmetry
breaking induced by an externally controlled time-dependent parameter
$\varepsilon(t)$. We consider linear quenches in $\varepsilon(t)$ with
functional dependence given by
\begin{align}
\label{eq:protocol}
\varepsilon(t) = \varepsilon_0
+\frac{t}{\tau_Q}\left(\varepsilon_1-\varepsilon_0 \right), \quad
\quad 0\leq t\leq \tau_Q.
\end{align}
where $\varepsilon(0)=\varepsilon_0>0$ and
$\varepsilon(\tau_Q)=\varepsilon_1<0$, so that the systems is in the
symmetric phase at the start of the quench protocol and in the
symmetry broken phase at the end of the quench protocol. The rate at
which the critical point is traversed is
$d{\varepsilon}(t)/dt=(\varepsilon_1-\varepsilon_0)/\tau_Q$, and thus,
it is determined by the quench time $\tau_Q$ once $\varepsilon_0$ and
$\varepsilon_1$ are fixed.

\begin{figure}
\centering
\includegraphics[width=1\linewidth,angle=00]{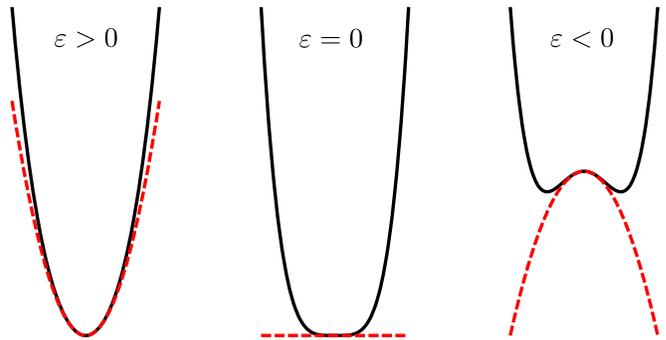}
\caption{{\small Schematic representation of the potential $V(\phi)=g\phi^4/4+\varepsilon\phi^2/2$ (solid black line) and its approximation $V(\phi)\approx \varepsilon \phi^2/2$ (dashed red line) for three different situations, $\varepsilon>0$ (left), $\varepsilon=\varepsilon_c=0$ (middle) and $\varepsilon<0$ (right).}}
\label{fig:1}
\end{figure}

{\em Langevin approach.---} The dynamics of the one-dimensional order
parameter $\phi(x,t)$ is described by the  following general
stochastic equation of
motion~\cite{Hohenberg:77,Chiara:10,Nigmatullin:16}
\begin{equation}
    \label{eq:stoch_eq}
    \left(\frac{\partial^2}{\partial t^2}+\eta\frac{\partial}{\partial
    t}\right)\phi(x,t)=h^2\frac{\partial^2}{\partial
    x^2}\phi(x,t)-\frac{\delta V(\phi)}{\delta \phi}+\zeta(x,t)
\end{equation}
where $\eta$ is the friction parameter and  $\zeta(x,t)$ the
stochastic force, which fulfills
\begin{align}
    \braket{\zeta(x,t)}&=0, \\ \braket{\zeta(x,t)\zeta(x',t')}
    &=\frac{2\eta}{\beta}\delta(x-x') \delta(t-t'),
\end{align}
where $\braket{\ldots}$ denotes the ensemble average, $\beta\equiv(k_BT)^{-1}$,
$T$ is the temperature and $k_B$ is the Boltzmann constant. For $\varepsilon\geq 0$,
the field $\phi(x)$, close to the ground state, has small amplitude such that
$|g\phi^3|\ll |\varepsilon\phi|$. Therefore,
to a good approximation, the higher order terms of $\phi$ in $V(\phi)$
can be neglected resulting in $V(\phi)\approx \varepsilon\phi^2/2$, as
depicted in Fig. \ref{fig:1}. Despite of this apparently naive
simplification,  the linearized stochastic equations of motion still
reproduce the dynamics of realistic
models~\cite{Moro:99,Chiara:10,Nigmatullin:16}. The linearized version
of Eq.~(\ref{eq:stoch_eq}) reads
\begin{equation}
\label{eq:stoch_eq2}
\left(\frac{\partial^2}{\partial t^2}+\eta\frac{\partial}{\partial
  t}\right)\phi(x,t)=\left(h^2\frac{\partial^2}{\partial
  x^2}-\varepsilon(t)\right)\phi(x,t)+\zeta(x,t).
\end{equation}
It is convenient to express the field $\phi(x,t)$ in the Fourier space
\begin{align}
\label{eq:modesF}
\phi(x,t)=\sum_{n} \varphi_n(t) e^{i k_n x},
\end{align}
where $k_n=2\pi n/L$.  For simplicity, we consider periodic boundary
conditions, $\phi(0,t)=\phi(L,t)$, and a real field $\phi(x,t)$, which
implies
$\varphi_{n}(t)=\varphi_{-n}^{*}(t)$. Substituting Eq.~(\ref{eq:modesF})
in Eq.~(\ref{eq:stoch_eq2}) results in decoupled equations of motion for
each mode. The equation for the $n$th mode reads
\begin{equation}
\label{eq:modes_dyn}
\left(\frac{\partial^2}{\partial t^2}+\eta\frac{\partial}{\partial
  t}\right)\varphi_n(t)=\left(-k_n^2
h^2-\varepsilon(t)\right)\varphi_n(t)+\zeta_n(t),
\end{equation}
with $\langle \zeta_n(t)\rangle=0$ and $\langle
\zeta_n(t)\zeta_m(t')\rangle=2\eta\beta^{-1}\delta_{nm}\delta(t-t')$.

As usual within the framework of statistical mechanics, we are
interested in the ensemble averages of observable macroscopic
quantities (e.g. correlation length) that are functions of the
microstates of the system. The ensemble averaged value
$\left<\mathcal{A}[\phi(t),\dot{\phi}(t);t] \right>$ of a physical
quantity $\mathcal{A}$ of interest at time $t$ may be obtained by
averaging over a number of stochastic trajectories generated by the
Langevin dynamics. Typically, the smaller the system, the more
stochastic trajectories are necessary to obtain a reliable and
meaningful average. Formally, the ensemble average can be obtained by
integrating over the field distribution expressed in the Fourier space
as
\begin{align}
\left<\mathcal{A}\right>=\prod \int_{-\infty}^{+\infty} d\varphi_n
\prod \int_{-\infty}^{+\infty}  d\dot{\varphi}_n \mathcal{A}  \prod
P_n(t,\varphi_n,\dot{\varphi}_n) ,
\end{align}
 where $P_n(t,\varphi_n,\dot{\varphi}_n)$ is the time-dependent
 probability distributions for $n$th mode in the phase space, i.e. it
 defines the probability of obtaining a value $\varphi_n$, and its
 velocity, $\dot{\varphi}_n$, at time $t$ for a mode with momentum
 $k_n=2\pi n/L$. The probability distribution is normalized according
 to
\begin{equation}
\int_{-\infty}^{+\infty} \int_{-\infty}^{+\infty} d\varphi_n
d\dot{\varphi}_n \, P_n(t,\varphi_n,\dot{\varphi}_n) = 1.
\end{equation}
In the Langevin approach, ensemble averaging involves evaluating
approximations to these probabilities from the repeated solutions of
the stochastic dynamical equations. The Fokker-Planck approach
provides an analytic expression for $P_n(t,\varphi_m,\dot{\varphi}_n)$
as a solution to fully deterministic partial differential equations,
as we explain in the following.

{\em Fokker-Planck approach.---} The Fokker-Planck formalism is a
well-known approach to handle stochastic dynamics, which, in contrast
to the Langevin approach focuses from the start on the probability
distributions of the stochastic variables. The dynamical equations for
the probability distributions are deterministic partial differential
equations~\cite{Risken:84}. The Fokker-Planck counterpart  of
Eq.~(\ref{eq:modes_dyn}) that specifies the dynamics of the $n$th
mode, reads
\begin{align}
\label{eq:kra}
&\frac{\partial P_n(t,\varphi_n,\dot{\varphi_n})}{\partial t} =
\left[-\frac{\partial}{\partial \varphi_n}
  \dot{\varphi}_n+\frac{\partial}{\partial \dot{\varphi}_n} \left(\eta
  \dot{\varphi}_n+\right.\right.\nonumber\\ &\qquad\left.\left.+\left(
  h^2k_n^2+\varepsilon(t)\right) \varphi_n\right)
  +\frac{\eta}{\beta}\frac{\partial^2}{\partial \dot{\varphi}_n^2}
  \right]P_n(t,\varphi_n,\dot{\varphi_n}),
\end{align}
which is known as the Kramers equation~\cite{Risken:84}. Hence the
full probabilistic dynamics is acquired solving Eq.~(\ref{eq:kra}).

{\em Quantities of interest.---} In the spirit of KZ mechanism, we
characterize the dynamics by means of the correlations induced in the
system as it traverses the second-order phase transition at
$\varepsilon_c=0$ at different rates
$d\varepsilon(t)/dt\propto\tau_Q^{-1}$.  To quantify such
correlations, we introduce the usual two-point correlation function
\begin{align}
\label{eq:2point}
G(x_1,x_2,t)=
\braket{\phi(x_1,t)\phi(x_2,t)}-\braket{\phi(x_1,t)}\braket{\phi(x_2,t)}.
\end{align}
As a consequence of the periodic boundary conditions
$\phi(0,t)=\phi(L,t)$, the two-point correlation function depends only
in the distance $x_1-x_2$, i.e. $G(x_1,x_2,t)\equiv G(x_1-x_2,t)$. The
correlation length of the field $\phi(x,t)$ can be defined as
\begin{align}
\label{eq:xi}
\xi_L(t)=\frac{\sqrt{\int_0^{L/2}dx\,  x^2 \,
    G(x,t)}}{\sqrt{2\int_0^{L/2} dx\, G(x,t) }}.
\end{align}
We also attempt to quantify the {\em defect} density formed during
the evolution, that is, the number of domains or regions per unit
length within $\phi(x,t)$ with an equivalent choice of the broken symmetry.
In the defect region the field interpolates rapidly but smoothly between
the chosen configurations and thus in those regions the field has large spatial
variations. For that reason the density of defects may be quantified by the
gradient of the field~\cite{Laguna:97}. Hence, to quantify such spatial
variations, we introduce the density $g_L$ as
\begin{align}
\label{eq:gL}
g_L(t)=L\frac{ \int_0^{L} dx \left< \left( \partial_x
  \phi(x,t)\right)^2\right>}{\int_0^{L} dx \left< \left(
  \phi(x,t)\right)^2\right>}.
\end{align}
The quantities $\xi_L(t)$ and $g_L(t)$  contain important non-equilibrium
dynamical information, allowing us to test the emergence of universal
behavior such as KZ scaling.

\emph{KZ scaling laws in the thermodynamic limit.---}   The equilibrium
correlation length in the thermodynamic limit diverges as $\xi\propto
\left|\varepsilon-\varepsilon_c\right|^{-\nu}$, where $\nu$ is the
corresponding critical exponent. Additionally, a second-order phase
transition is also characterized by a diverging relaxation time,
$\tau\propto\left|\varepsilon-\varepsilon_c\right|^{-\nu z}$, where $z$ is
the dynamical critical exponent~\cite{Hohenberg:77}. If the fourth and
higher order terms of $\phi$ in Eq.~(\ref{eq:GLpot}) are negligible then
$\nu=1/2$ (mean field exponent), while depending on the dynamical regime,
$z=2$ or $z=1$, for overdamped or underdamped
dynamics~\cite{Laguna:98,Chiara:10}. The former regime is found when
$\left|\eta \partial_t \phi \right|\gg \left|\partial^2_{t^2} \phi\right|$,
while the latter takes place in the opposite limit. To derive scaling laws,
we resort to the KZ argument, which states that due to the diverging
relaxation time near the critical point, there will be a \emph{freeze-out}
instance, $\hat{t}$, at which the system is not able any longer to adjust its
correlation length to its equilibrium value~\cite{Kibble:76,delCampo:14}
during the quench. Accordingly, traversing the critical point at a
finite rate $\tau_Q^{-1}$ provokes the formation of defects or excitations,
whose typical size scales as $\hat{\xi}\sim \tau_Q^{\nu/(1+z\nu)}$, where
$\hat{\xi}$ corresponds to the correlation length at the {\em freeze-out}
instant. Therefore, the density of excitations will scale as
$L^d/\hat{\xi}^d\sim\tau_Q^{-d\nu/(1+z\nu)}$, where $d$ is the dimension of
the system. The KZ scaling laws can also be derived using rescaling
transformations of the equations of motion~\cite{Gor:16}, which does not rely
on the physical arguments of transition between adiabatic and impulsive
dynamics~\cite{Kibble:76,delCampo:14}. In this paper, we show how these
rescaling transformations can be applied to our system, elucidating a set of
non-equilibrium scaling functions, where KZ scaling laws appear as a special
case.

\section{Overdamped regime: Smoluchowski equation}
\label{sec:smo}
The general equation of motion which governs the dynamics is given in
Eqs.~(\ref{eq:stoch_eq2}) and~(\ref{eq:kra}) for Langevin and Fokker-Planck
formalism, respectively. However, when the term $\eta \partial_t \phi$
dominates, $|\eta \partial_t \phi|\gg |\partial^2_{t^2}\phi|$, the dynamics
of $\phi(x,t)$ is overdamped or pure relaxational~\cite{Hohenberg:77}. In the
overdamped regime, the Langevin equation of motion reads
\begin{equation}
    \eta\frac{\partial }{\partial t}\phi(x,t)=\left(h^2\frac{\partial^2}{\partial x^2}-\varepsilon(t)\right)\phi(x,t)+\zeta(x,t).
\end{equation}
The Fourier decomposition Eq.~(\ref{eq:modesF}) results in decoupled
equations for each of the normal modes. The dynamical equation for the $n$th
mode is
\begin{equation}
    \label{eq:overdamped_langevin}
    \eta\frac{\partial }{\partial t}\varphi_n(t)=\left(-h^2 k_n^2-\varepsilon(t)\right)\varphi_n(t)+\zeta_n(t),
\end{equation}
with $\langle \zeta_n(t)\rangle=0$ and $\langle \zeta_n(t)
\zeta_m(t')\rangle=2\eta \beta^{-1} \delta_{nm}\delta(t-t') $.
The Fokker-Planck equation in the overdamped regime is known as the
Smoluchowski equation. Let $P_{o,n}(t,\varphi_n)$ denote the probability
distribution for the $n$th mode with $k_n=2\pi n/L$; the subscript $o$ emphasizes the overdamped
nature of the dynamics. Since $\varphi_n(t)$ is in general complex, it is
more convenient to express the field as
\begin{align}
    \label{eq:field_ri}
    &  \qquad\phi(x,t)=\frac{\varphi_0^R(t)}{\sqrt{N_c}}\nonumber\\&
    +\sqrt{\frac{2}{N_c}}\sum_{n=1}^{(N_c-1)/2}\varphi_n^R(t)\cos(k_nx)+\varphi_n^I(t)\sin(k_nx),
\end{align}
where we have introduced a momentum cut-off $k_{c}$ which sets a maximum number of modes $(N_c-1)/2$, and $\varphi_n^R(t)\equiv\textrm{Re}(\varphi_n(t))$ and
$\varphi_n^I(t)\equiv\textrm{Im}(\varphi_n(t))$ due to the condition
$\varphi_{-n}(t)=\varphi_n^*(t)$. Note that, without loss of generality,
$N_c$ is chosen to be odd. Then, the Smoluchowski equation for
$P_{o,n}(t,\varphi_n^{R,I})$ reads~\cite{Risken:84}
\begin{align}
    \label{eq:smo}
    &\eta \frac{\partial P_{o,n}(t,\varphi_n^{R,I})}{\partial t}= \frac{\partial }{\partial \varphi_n^{R,I}}\left[\frac{1}{\beta} \frac{\partial }{\partial \varphi_n^{R,I}}+\right.\nonumber\\&\qquad+\left.\left(h^2k_n^2+\varepsilon(t) \right)\varphi_n^{R,I}\right]P_{o,n}(t,\varphi_n^{R,I}).
\end{align}
 We will assume that the system is initially in thermal equilibrium at
$\varepsilon_0$. In thermal equilibrium, a probability distribution,
$P^{\rm{th}}_{o,n}$, must fulfill
\begin{equation}
    \label{eq:Pthermal}
    \partial_t P^{\rm th}_{o,n}=0.
\end{equation}
Substituting Eq.~(\ref{eq:Pthermal}) in~(\ref{eq:smo}) and solving the resulting differential equation gives
\begin{align}
\label{eq:th_smo}
    P_{o,n}^{\textrm{th}}(\varphi_n^{R,I})&= \frac{f_n^{\rm th}}{\sqrt{\pi}} e^{-(f_n^{\rm th}\varphi_n^{R,I})^2},\nonumber \\
    f_{n}^{\rm th}&=\sqrt{\beta\left( h^2k_n^2+\varepsilon_0\right)/2}.
\end{align}
As expected, these probabilities correspond to the Boltzmann distribution at
given temperature $1/\beta$ of the bath. Note that they exist only for
$\varepsilon_0>\varepsilon_c=0$ as a consequence of the harmonic
approximation $V(\phi)\approx \varepsilon\phi^2/2$. Having determined the
initial state, we now solve Eq.~(\ref{eq:smo}) to find the time-dependent
probability distribution $P_{o}(t)$. For that, we make use of a Gaussian
Ansatz
\begin{align}
\label{eq:ansatz_smo}
    P_{o,n}(t)=\frac{f_{n}(t)}{\sqrt{\pi}} e^{-f_n^2(t)\varphi_n^2}.
\end{align}
Substituting Eq.~(\ref{eq:ansatz_smo}) into Eq.~(\ref{eq:smo}) gives
\begin{align}
\label{eq:difeqover}
    \frac{\partial}{\partial t} f_n(t)&= -\frac{2}{{\eta \beta}} f_n^3(t)+\frac{1}{\eta}\left(h^2k_n^2+\varepsilon(t) \right)  f_n(t),
\end{align}
with the initial condition determined by the thermal equilibrium,
$f_n(0)=f_n^{\rm th}$. Thus the full knowledge of the probabilistic
dynamics is captured in a set of uncoupled differential equations  for the
variance of each probability distribution.

The knowledge of the functional form of the probability distributions, allows
us to explicitly calculate the quantities of interests, namely $\xi_L(t)$ and
$g_L(t)$. Substituting Eq.~(\ref{eq:ansatz_smo}) into the expression for the
two-point correlation function given by Eq.~(\ref{eq:2point}), and
simplifying the resulting expression gives
\begin{align}
\label{eq:2p_smo}
    G(x_1,x_2,t)=\frac{1}{2N_cf_0^2(t)}+\sum_{n=1}^{N_c-1}\frac{\cos(k_n(x_1-x_2))}{N_cf_n^2(t)}.
\end{align}
The correlation length is then obtained using Eq.~(\ref{eq:xi}),
\begin{align}
    \label{eq:xi_smo}
    &\xi_L(t) =\frac{L}{2\sqrt{6}}\sqrt{1+12f_0^2(t)\sum_{n=1}^{(N_c-1)/2}\frac{(-1)^n}{f_n^2(t)n^2\pi^2}}.
\end{align}
Similarly, the expression for $g_L(t)$ evaluates to
\begin{align}
    \label{eq:gL_smo}
    g_L(t)=L\frac{\sum_{n=1}^{(N_c-1)/2}\frac{k_n^2}{f_n^2(t)}}{\frac{1}{2f_0^2(t)}+\sum_{n=1}^{(N_c-1)/2}\frac{1}{f_n^2(t)}}.
\end{align}
Thus, we have obtained analytic expressions for the correlation length and
the number of defects as a function of time for non-equilibrium Kibble-Zurek
dynamical scenario. In the analytical expressions, we can now look for
physical meaning such as the presence of non-equilibrium scaling laws. We
break down the discussion into three parts, each of which corresponds to a
different quench rate regime: infinitely slow or isothermal quench (the case
of thermal equilibrium), sudden quench and finite-rate quench.

{\em Thermal equilibrium.---} In the limit $\tau_Q\rightarrow \infty$,
thermal equilibrium is achieved at any {$\varepsilon(t)$}. Although this
scenario is just a limiting case of a finite-rate quench, it allows us to
gather some interesting equilibrium properties which will be helpful later
on.  We stress that the derived expressions given in Eqs.~(\ref{eq:xi_smo})
and~(\ref{eq:gL_smo}) are valid for either  equilibrium or non-equilibrium.
Equilibrium properties are recovered simply by considering thermal
probability distributions $P_{o,n}^{\rm th}$ at any $\varepsilon$, that is,
$f_{n}^{\rm th}$, given in Eq.~(\ref{eq:th_smo}). In the limit
$N_c \rightarrow \infty$, the correlation length at thermal equilibrium for
finite $L$ reads
\begin{align}
    \label{eq:xi_th}
    \xi^{\textrm{th}}_L(\varepsilon>0)=\sqrt{\frac{h^2}{\varepsilon}-\frac{hL}{2\varepsilon^{1/2}\sinh(\varepsilon^{1/2}L/2h)}},
\end{align}
while in the thermodynamic limit, $L,N_c\rightarrow\infty$, keeping the
cut-off $k_c$ finite, is just $\xi^{\rm th}_{L\rightarrow\infty}(\varepsilon>0)
= h\sqrt{1/\varepsilon}$. As we expected for Ginzburg-Landau theory, we
obtain a  critical exponent $\nu=1/2$ for the diverging correlation length
at the critical point, i.e., $\xi\propto|\varepsilon-\varepsilon_c|^{-\nu}$.
At $\varepsilon=\varepsilon_c=0$, the resulting expression is particularly
simple for finite $L$,
\begin{align}
    \label{eq:xi_th_cr}
    \xi^{\textrm{th}}_L(\varepsilon_c=0)&=\frac{L}{2\sqrt{6}}.
\end{align}
As one expects for a finite system, the correlation length can not exceed the
system size, reaching its maximum at the critical point $\xi_L\lesssim L$.
The previous result gives precisely its saturation value in the harmonic
approximation of the Ginzburg-Landau model, as well as the scaling
$\xi_L(\varepsilon_c=0)\propto L$ in agreement with finite-size scaling
theory~\cite{Fisher:72,Hohenberg:77}.

In a similar way, we can calculate the $g_L^{\textrm{th}}(\varepsilon)$ in the thermodynamic limit as
\begin{align}
g_{L\rightarrow\infty}^{\rm th}(\varepsilon>0)/L&\approx \frac{\int_0^{k_c}dk \frac{k^2}{\beta \left(h^2k^2+\varepsilon \right)} }{\int_0^{k_c}dk \frac{1}{\beta \left(h^2k^2+\varepsilon \right)}}\nonumber \\&=\frac{\varepsilon^{1/2}k_c}{h\arctan({hk_c\varepsilon^{-1/2}})}-\frac{\varepsilon}{h^2},
\end{align}
and hence, as $\varepsilon\rightarrow \varepsilon_c=0$, it vanishes as
\begin{align}
  g_{L\rightarrow\infty}^{\rm th}/L\sim\frac{2k_c}{h\pi}(\varepsilon-\varepsilon_c)^{1/2},
\end{align}
revealing its critical exponent, which turns out to be $1/2$.

{\em Sudden quench limit.---} We briefly comment on the limit of sudden
quenches, that is, when $\tau_Q\rightarrow 0$. In this case, as the system
has no time to react to external perturbations, the corresponding properties
of the system remain unchanged from its initial thermal state. Therefore, the
results of sudden quenches are simply given by the thermal initial state at
$\varepsilon_0$, i.e., $\xi_L^{\rm th}(\varepsilon_0)$ and $g_L^{\rm
th}(\varepsilon_0)$, which for the former the expression is explicitly given
in Eq.~(\ref{eq:xi_th}).
\begin{figure}
\centering
\includegraphics[width=0.51\linewidth,angle=-90]{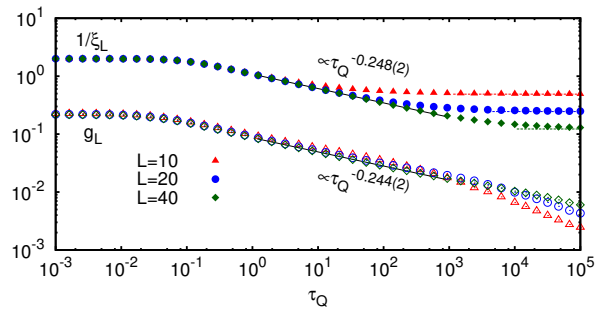}
\caption{{\small Results for $\xi_L$ and $g_L$ in the overdamped regime as a
function of the quench time $\tau_Q$ for three different system sizes,
$L=10$, $20$ and $40$ with a fixed cut-off momentum $k_c=5\pi$ right at the
critical point. Note that, for a better visualization, $g_L$ is divided by a
$40$. The results clearly show a power-law scaling
$\tau_Q^{-\nu/(1+z\nu)}=\tau_Q^{-1/4}$ for intermediate quench rates as
predicted by KZ mechanism. The solid black lines display the fit to a
power-law for $L=40$, together with the resulting exponent for both
quantities.  Dashed lines correspond to the minimum value of $1/\xi_L$, i.e.,
$2\sqrt{6}/L$ for the three different system sizes. Results obtained with
$h=5$, $\beta=1$, $\eta=10$, $\varepsilon_0=100$ and $\varepsilon_1=-10$.}}
\label{fig:2}
\end{figure}
\begin{figure}
\centering
\includegraphics[width=0.51\linewidth,angle=-90]{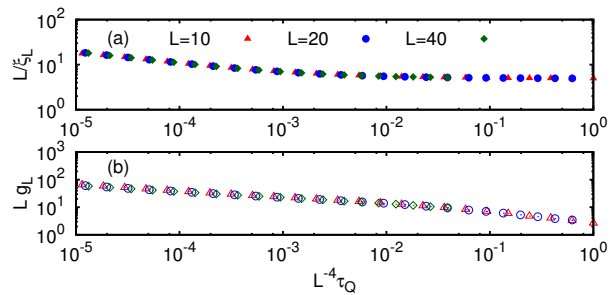}
\caption{{\small Non-equilibrium finite-size  scaling functions in the
overdamped regime. The data collapse for different system sizes and quench
times for (a) $1/\xi_L$ and (b) $g_L$  confirm the relation $1/\xi_L\sim
\tau_Q^{-\nu/(1+z\nu)} y_{\xi}(\tau_Q L^{-1/\nu-z})$ and $g_L\sim
\tau_Q^{-\nu/(1+z\nu)}y_{g}(\tau_Q L^{-1/\nu-z})$, respectively, with
$\nu=1/2$ and $z=2$. }}
\label{fig:3}
\end{figure}
\begin{figure}
\centering
\includegraphics[width=0.51\linewidth,angle=-90]{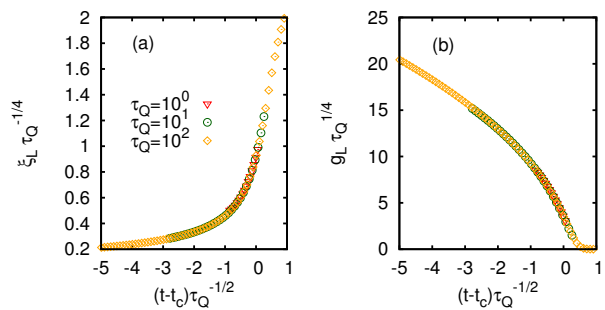}
\caption{{\small Collapse of (a) $\xi_L$ and (b) $g_L$ for $L=40$ into a
single curve during the whole evolution for the overdamped regime and for
different quench times $\tau_Q$.}}
\label{fig:4}
\end{figure}

{\em Finite-rate quenches.---} We consider now the non-equilibrium dynamics
for finite $\tau_Q$ in the overdamped regime, where KZ theory predicts
scaling laws as a function of the quench rate. That is, we quench linearly
the parameter $\varepsilon(t)$ in a time $\tau_Q$ according to
Eq.~(\ref{eq:protocol}), and solve the equations of
motion~(\ref{eq:difeqover}), by numerical integration.  Numerical solutions
can be easily done by means of standard Runge-Kutta techniques. We emphasize
that solving Eq.~(\ref{eq:difeqover}) immediately allows us to calculate
precise average quantities, while the Langevin approach requires evaluating
many realizations, which is far more costly from a computational point of
view.

We set a momentum cut-off of $k_{N_c}=5\pi$, which leads to a maximum number
of modes $N_c$ for a given length $L$. Initially, the system is prepared in
thermal equilibrium at $\varepsilon_0=100$ and quenched in a time $\tau_Q$
towards $\varepsilon_1=-10$. The other parameters are set to $h=5$, $\eta=10$
and $\beta=1$. The results for three different system sizes at
$\varepsilon(t)=0$, $L=10$, $20$ and $40$, are presented in Fig.~\ref{fig:2},
where $g_L$ and $1/\xi_L$ exhibit KZ scaling at intermediate quench rates. As
the system size increases, the region of universal power-law scaling gets
broader. For very fast quenches, $\tau_Q\rightarrow 0$, $1/\xi_L$ and $g_L$
saturate to their initial value, while for $\tau_Q\rightarrow\infty$, they
tend to its value at thermal equilibrium, $g_L\rightarrow 0$ and
$1/\xi_L\rightarrow 2\sqrt{6}/L$, as explained previously.  For $L=40$ we
perform a fit to obtain the power-law exponent which agrees well with the KZ
prediction, $\tau_Q^{-\nu/(1+z\nu)}=\tau_Q^{-1/4}$ for the overdamped regime.
Furthermore, we illustrate the finite-size scaling at intermediate quench
rates, which predicts that
\begin{eqnarray}
    1/\xi_L&\sim& \tau_Q^{-\nu/(1+z\nu)}y_\xi(\tau_Q L^{-1/\nu-z})\\
    g_L&\sim& \tau_Q^{-\nu/(1+z\nu)}y_g(\tau_Q L^{-1/\nu-z}),
\end{eqnarray}
 where $y_{\xi}(x)$ and $y_g(x)$ are non-equilibrium scaling functions which
 fulfill $y(x\ll 1)\sim$ constant (where KZ scaling law emerges) and $y(x\gg
 1)\sim x^{\nu/(1+z\nu)}$, for nearly adiabatic
 quenches~\cite{Nigmatullin:16}. For that, we plot $L/\xi_L$ which is
 expected to follow a functional form $L\tau_Q^{-\nu/(1+z\nu)}y_\xi(\tau_Q
 L^{-1/\nu-z})$ or simply $x^{-\nu/(1+z\nu)}y_{\xi}(x)$ being $x=\tau_Q
 L^{-1/\nu-z}$. Thus, $L/\xi_L$ and $g_L L$ depend only on the scaling
 variable $\tau_Q L^{-1/\nu-z}$. The collapse of the data onto a single curve
 shown in Fig.~\ref{fig:3} confirms this non-equilibrium scaling hypothesis.

Additionally, we demonstrate the universality of the phase transition
dynamics by transforming physical quantities in such a way that any
$\tau_Q$-dependence is removed from the equations of motion, following the
theory developed in~\cite{Gor:16}. This is achieved by
performing transformations $x\rightarrow x\tau_Q^{-1/4}$ and $t\rightarrow
(t-t_c)\tau_Q^{-1/2}$, where $t_c=\tau_Q\frac{\varepsilon_0}{\varepsilon_0 - \varepsilon_1}$ is the
 instance at which the critical point is crossed. In this rescaled frame the
 dynamics is universal and hence the functional dependence of $\xi_L
 \tau_Q^{-1/4}$ and $g_L \tau_Q^{1/4}$ on $(t-t_c)\tau_Q^{-1/2}$ is expected
 to be the same irrespective of the value of $\tau_Q$. This universality
 during the whole evolution is demonstrated in Fig.~\ref{fig:4}, which shows
 the collapse of the results for three different values of $\tau_Q$ onto two
 curves, one for $\xi_L$ and one for $g_L$.


\section{General and underdamped regime: Kramers equation}
\label{sec:kram}
Let us recall that the Kramers equation Eq.~(\ref{eq:kra}) describes the
general dynamical regime that includes both dissipative and inertial terms.
In our particular case, the Fokker-Planck equation which describes the
dynamics reads
\begin{align}
    \label{eq:Kramer}
    &\frac{\partial P_n(t,\varphi_n,\dot{\varphi}_n)}{\partial t} = \left[-\frac{\partial}{\partial \varphi_n} \dot{\varphi}_n+\frac{\partial}{\partial \dot{\varphi}_n} \left(\eta \dot{\varphi}_n+\right.\right.\nonumber\\
    &\qquad\left.\left.+\left( h^2k_n^2+\varepsilon(t)\right) \varphi_n\right) +\frac{\eta}{\beta}\frac{\partial^2}{\partial \dot{\varphi}_n^2} \right]P_n(t,\varphi_n,\dot{\varphi}_n),
\end{align}
where $P_n(t,\varphi_n(t),\dot{\varphi}_n(t))$ is now a two-dimensional
probability distribution at time $t$. The analysis of Eq.~(\ref{eq:Kramer})
is more intricate, but nevertheless the procedure is similar  to the one
presented in Sec.~\ref{sec:smo} for the overdamped dynamics.

Thermal equilibrium states are obtained from  $\partial_t P_n^{\textrm{th}}=0$, whose solution in terms of $\varphi_n^{R}\equiv{\rm Re}(\varphi_n)$ and $\varphi_n^{I}\equiv{\rm Im}(\varphi_n)$ reads
\begin{align}
\label{eq:th_kram}
    P_{n}^{\textrm{th}}&=\frac{\sqrt{(A_n^{\textrm{th}}B_n^{\textrm{th}})^2-(C_n^{\textrm{th}})^2}}{2\pi} \, \times\nonumber \\
    \times \textrm{Exp}&\left[-\frac{1}{2}\left(\left(A_n^{\textrm{th}}\varphi_n^{R,I}\right)^2+\left(B_n^{\textrm{th}}\dot{\varphi}_n^{R,I}\right)^2-2C_n^{\textrm{th}}\varphi_n^{R,I} \dot{\varphi}_n^{R,I}\right)\right]
\end{align}
where
\begin{align}
    A_{n}^{\textrm{th}}&= \sqrt{\beta\left(h^2k_n^2+\varepsilon\right)},\\
    B_{n}^{\textrm{th}}&= \sqrt{\beta}, \\
    C_{n}^{\textrm{th}}&=0.
\end{align}
The time evolution of the probability distributions $P_n(t)$ is given by time-dependent coefficients $A_n(t)$, $B_n(t)$ and $C_n(t)$.  In this way, three coupled differential equations per mode under the protocol $\varepsilon(t)$ determine the dynamics,
\begin{align}
\label{eq:difeqkra}
\frac{\partial A_n(t)}{\partial t} &= -\frac{C_n(t)}{A_n(t)} \left( \varepsilon(t)  + h^2k_n^2 +\frac{\eta C_n(t)}{\beta} \right), \\
\frac{\partial B_n(t)}{\partial t} &= \eta B_n(t)-\frac{\eta}{\beta}B_n^3(t) +\frac{C_n(t)}{B_n(t)}, \\
\frac{\partial C_n(t)}{\partial t} &= A_n^2(t)+\eta C_n(t)-B_n^2(t)\left( h^2k_n^2+\right. \nonumber\\ &\hspace{3.0cm}\left.+\varepsilon(t) +\frac{2\eta C_n(t)}{\beta}\right).
\end{align}
 The average quantities are obtained in the same way as in the overdamped regime, but with the time-dependent probability distributions also dependent on $\dot{\varphi}_n$. Indeed, we can define the probability distribution $Q_n(t,\varphi_n^{R,I})$ once the velocity dependence is integrated out,
\begin{align}
Q_n(t,\varphi_n^{R,I})&=\int_{-\infty}^{+\infty}  d\dot{\varphi}_{n}^{R,I}\, P_n(t,\varphi_n^{R,I},\dot{\varphi}_n^{R,I})\nonumber \\
&=\frac{F_n(t)}{\sqrt{\pi}}e^{-F_n^2(t) (\varphi_n^{R,I})^2},
\end{align}
where $F_n(t)$ depends on the coefficients $A_n(t)$, $B_n(t)$ and $C_n(t)$ as
\begin{align}
\label{eq:F_def}
 F_n(t) &= \sqrt{\frac{1}{2B_n^2(t)}\left(A_n^2(t)B_n^2(t)-C_n^2(t) \right)}.
\end{align}
This allows us to directly apply the same expressions as those derived for the overdamped regime. Eqs.~(\ref{eq:xi_smo}) and~(\ref{eq:gL_smo})  for correlation length $\xi_L(t)$ and density $g_L(t)$ can be directly applied by just replacing $f_n(t)$ with $F_n(t)$.

{\em Thermal equilibrium and sudden quenches.---} Clearly, thermal
equilibrium does not depend on the considered dynamical regime. Therefore,
the same thermal equilibrium probability distributions are retrieved from
Eq.~(\ref{eq:th_kram}) and we refer to Sec.~\ref{sec:smo} for the
discussion on equilibrium features, as well as the opposite limit,
$\tau_Q\rightarrow 0$, of sudden quenches.
%
\begin{figure}
\centering
\includegraphics[width=0.51\linewidth,angle=-90]{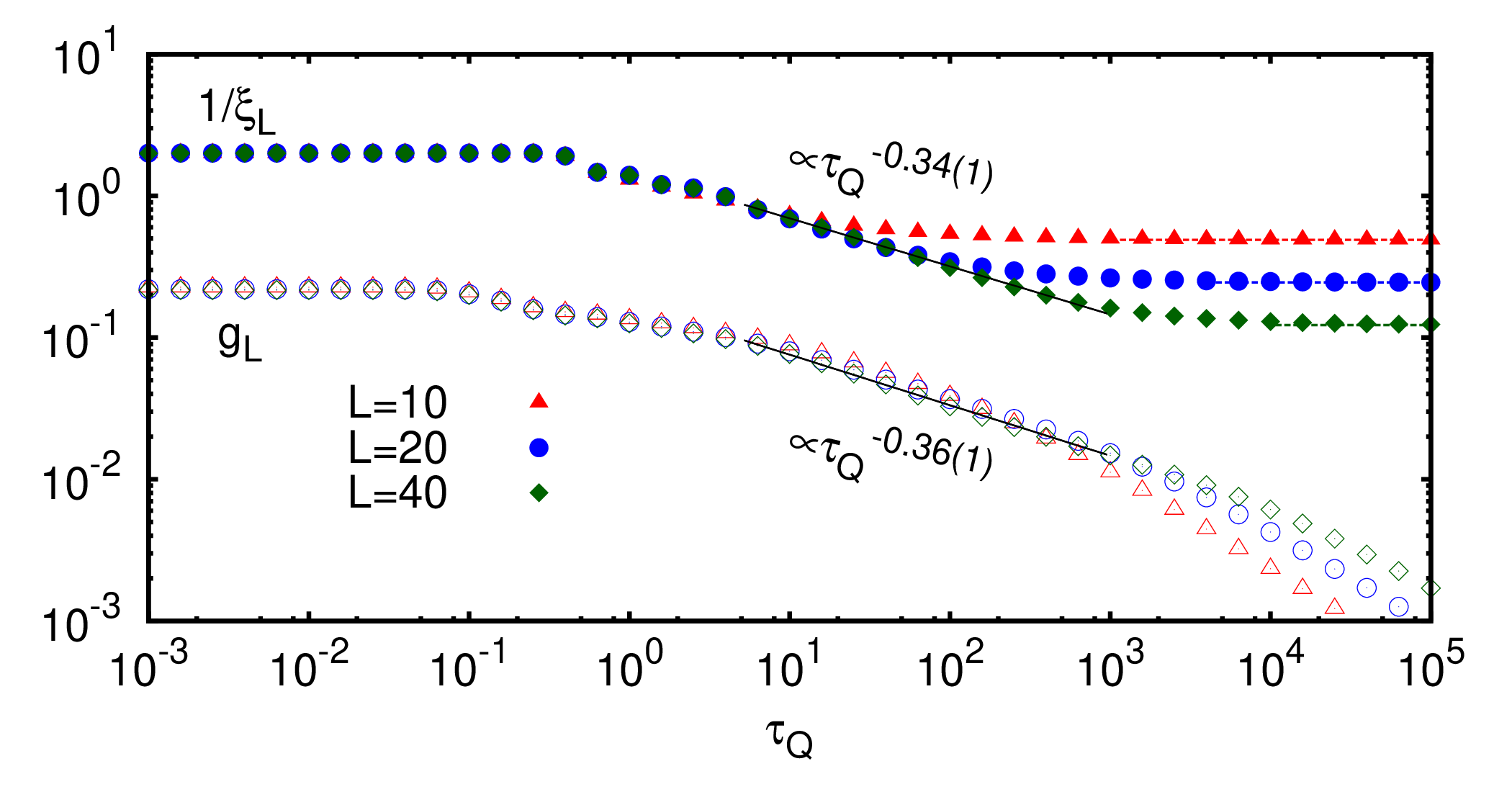}
\caption{{\small Results for $\xi_L$ and $g_L$  in the underdamped regime
($\eta=0.1$) as a function of the quench time $\tau_Q$ for three different
system sizes, $L=10$, $20$ and $40$ with a fixed cut-off momentum $k_c=5\pi$
right at the critical point. Note that $g_L$ is divided by $40$ for a better
visualization. The results clearly show a power-law scaling
$\tau_Q^{-\nu/(1+z\nu)}=\tau_Q^{-1/3}$ for intermediate quench rates as
predicted by the theory. The solid black lines display the fit to a power-law
for $L=40$, together with the resulting exponent.  Dashed lines correspond to
the minimum value of $1/\xi_L$, i.e., $2\sqrt{6}/L$ for the three different
system sizes. Results were obtained with $h=5$, $\beta=1$,
$\varepsilon_0=100$ and $\varepsilon_1=-10$.}}
\label{fig:5}
\end{figure}
\begin{figure}
\centering
\includegraphics[width=0.51\linewidth,angle=-90]{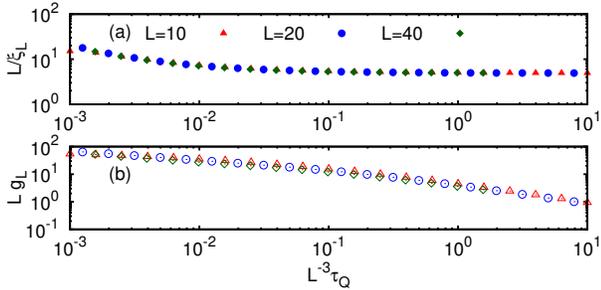}
\caption{{\small Non-equilibrium finite-size scaling functions in the
underdamped regime ($\eta=0.1$) for different system sizes a quench times.
The data collapse for (a) $1/\xi_L$ and (b) $g_L$ confirm the relation
$1/\xi_L\sim \tau_Q^{-\nu/(1+z\nu)} y_{\xi}(\tau_Q L^{-1/\nu-z})$ and
$g_L\sim \tau_Q^{-\nu/(1+z\nu)}y_{g}(\tau_Q L^{-1/\nu-z})$, respectively,
being the critical exponents in the underdamped case $\nu=1/2$ and $z=1$.}}
\label{fig:6}
\end{figure}
\begin{figure}
\centering
\includegraphics[width=0.51\linewidth,angle=-90]{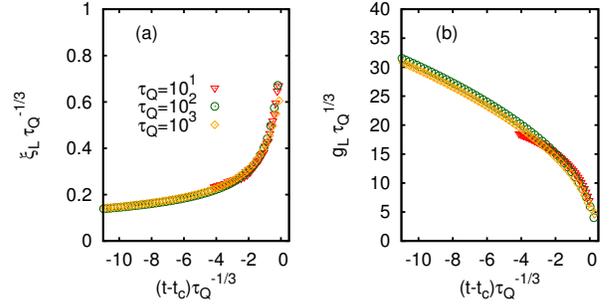}
\caption{{\small Collapse of (a) $\xi_L$ and (b) $g_L$ for $L=40$ into a
single curve during the whole evolution for the underdamped regime and for
different quench times $\tau_Q$. }}
\label{fig:7}
\end{figure}

{\em Finite-rate quenches.---} As in the case of overdamped dynamics, the KZ
scaling laws are observed at finite quench rates. We solve numerically the
equations of motion Eq.~(\ref{eq:difeqkra}) to obtain the time-dependent
probability distributions for different $\tau_Q$. Then, from $F_n(t)$ we
calculate $\xi_L(t)$ and $g_L(t)$ using Eqs.~(\ref{eq:xi_smo})
and~(\ref{eq:gL_smo}), respectively.
As in the overdamped regime, we set a momentum cut-off of $k_c=5\pi$. The
initial thermal state at $\varepsilon_0=100$ is quenched in a time $\tau_Q$
towards $\varepsilon_1=-10$. Additionally, we set $h=5$ and $\beta=1$. To
illustrate the KZ scaling in the underdamped regime, we select a small
friction coefficient $\eta=0.1$. The results are presented in
Fig.~\ref{fig:5} for three different system sizes, $L=10$, $20$ and $40$,
where the latter already exhibits a power-law scaling $\sim \tau_Q^{\alpha}$
for wide range of quench times. The performed fit gives an exponent
$\alpha=-0.34(1)$ for $1/\xi_L$, in agreement with the predicted KZ scaling
$\tau_Q^{-\nu/(1+z\nu)}=\tau_Q^{-1/3}$ since $\nu=1/2$ and $z=1$. A more
pronounced deviation is found for $g_L$, with $\alpha=-0.36(1)$, which might
be caused by finite-size effects. Moreover, in Fig.~\ref{fig:6} the
finite-size scaling at intermediate quench rates is verified. The data
collapse onto a single curve corroborates the relations  $1/\xi_L\sim
\tau_Q^{-\nu/(1+z\nu)}y_\xi(\tau_Q L^{-1/\nu-z})$ and $g_L\sim
\tau_Q^{-\nu/(1+z\nu)}y_g(\tau_Q L^{-1/\nu-z})$. Recall that $L/\xi_L$ and $L
g_L$ are expected to follow $x^{-\nu/(1+z\nu)} y_\xi(x)$ and
$x^{-\nu/(1+z\nu)} y_g(x)$ where $x=\tau_Q L^{-1/\nu-z}$ is a scaling
variable and $y(x)$ are non-equilibrium scaling functions.

Finally, we exemplify the  universality of the dynamics in the underdamped
regime. If  $\left|\eta \partial_t \phi \right|\ll \left|\partial^2_{t^2}
\phi\right|$, one removes the $\tau_Q$ dependence by performing the
transformation $x\rightarrow x\tau_Q^{-1/3}$ and $t\rightarrow
(t-t_c)\tau_Q^{-1/3}$~\cite{Gor:16}. Note that the transformation is
different to the overdamped case. The quantities $\xi_L \tau_Q^{-1/3}$ and
$g_L\tau_Q^{1/3}$ are expected to collapse for different quench times
$\tau_Q$ when plotted against the rescaled time $(t-t_c)\tau_Q^{-1/3}$. This
collapse is shown in Fig.~\ref{fig:7}, where we plot the results of the
calculations of $\xi_L(t)$ and $g_L(t)$ for three different values of
$\tau_Q$ in the rescaled coordinates.

\section{Coulomb crystals: linear to zigzag phase transition}
\label{sec:crys}

The analysis in the previous sections has been done for phase transitions
described by a one-dimensional Ginzburg-Landau field theory. However, the
Fokker-Planck approach is valid beyond the Ginzburg-Landau theory; the
knowledge of the quench function and the dispersion relation of the system is
sufficient to predict the expected density of defects or any other statistical
observable. We will illustrate this by applying our method to the problem of
dynamic structural phase transition in Coulomb
crystals~\cite{Fishman:08,Chiara:10}.

Coulomb crystals are ordered structures that form when charged particles
in a global confining potential are cooled below a critical temperature.
An example of the physical realization of Coulomb crystals are ion crystals
in Paul traps. Structural transitions in Coulomb crystal can be induced
by varying the global confining potential~\cite{Retzker:08,Partner:15}. The KZ mechanism of defect
formation was studied numerically and experimentally using linear
to zigzag non-equilibrium phase transition in ion traps~\cite{Pyka:13,Ulm:13}. The analysis
in references~\cite{delCampo:10,Chiara:10,Nigmatullin:16} relied on the
 mapping of the linear to zigzag transition
to a Ginzburg-Landau field theory model. In this section, we show
how to use the methods developed in this paper to analyze the dynamic
linear to zigzag transition without resorting to Ginzburg-Landau theory.

We consider $N$ charged particles moving in a periodic cell of size
$L$. The periodic boundary conditions simplify the analysis since they result in a homogeneous Coulomb crystal. Moreover, periodic boundary condition can be realized with the existing technology of ring ion traps~\cite{Wang:15, Li:16}. The potential energy of the $N$ particles reads
\begin{equation}
V=\frac{1}{2}\sum_{j=1}^{N}m\omega_{t}^{2}z_{j}^{2}+\frac{1}{2}\sum_{j=1}^{N}\sum_{j\neq i}\frac{Q^{2}}{|\textbf{r}_{i}-\textbf{r}_{j}|},\label{eq:Vion}
\end{equation}
where $\textbf{r}_{i}=(x_{j},z_{j})$ is the coordinate of the $j$th
ion, $m$ is the mass of the ions, $\omega_{t}$ is the transverse
trapping secular frequency and $Q^{2}\equiv e^{2}/4\pi\epsilon_{0}$.
There exists a critical frequency value $\omega_{t}^{c}=\sqrt{7\zeta_R(3)/2}\omega_0$,
where $\omega_0=\sqrt{Q^2/ma^3}$ and $\zeta_R(x)$ is the Riemann zeta function.
For $\omega_{t}>\omega_{t}^{c}$ the lowest energy configuration is a linear chain
and for $\omega_{t}<\omega_{t}^{c}$ the lowest energy configuration is a two-row zigzag chain.

Initially, the $N$ ions are in thermal equilibrium in a linear chain configuration. The transverse frequency $\omega_{t}$ is then quenched linearly in time through the critical point $\omega_t^c$, thereby inducing a transition from a linear to zigzag configuration at a rate proportional to $1/\tau_Q$
\begin{equation}
\omega_{t}(t)=\begin{cases}
\omega_{i} & \quad\textrm{for }t<0\\
\omega_{i}+ \frac{\omega_f-\omega_i}{\tau_Q}t & \quad\textrm{for }0\leq t<\tau_Q\\
\omega_{f} & \quad\textrm{for t}\geq\tau_{Q},
\end{cases}\label{eq:ion_quench}
\end{equation}
where $\omega_{i}>\omega_{t}^{c}$ and $\omega_{f}<\omega_{t}^{c}$.
Since the quench is performed at finite rate, the system is driven
out of equilibrium and there is a non-zero probability of formation
of a number of structural defects.

\emph{Langevin approach.---} The expectation of any observable $\left\langle \mathcal{A}(t)\right\rangle $
(including the number of defects) can be evaluated by repeatedly solving
the stochastic equations of motion that describe the dynamics of the
systems, and then estimating the expectation from the obtained sample
of trajectories. The dynamics of the system is determined by the following Langevin equations
of motion
\begin{eqnarray}
m\frac{d^{2}x_{j}}{dt^{2}}+\eta\frac{dx_{j}}{dt}+\frac{\partial V}{\partial x_{j}} & = & \zeta_{j}^{x}(t),\label{eq:ion_motion}\\
m\frac{d^{2}z_{j}}{dt^{2}}+\eta\frac{dz_{j}}{dt}+\frac{\partial V}{\partial z_{j}} & = & \zeta_{j}^{z}(t).\label{eq:ion_motion2}
\end{eqnarray}
There, $\eta$ is the friction coefficient and $\zeta_{j}^{x,z}$ is
the stochastic force that satisfies the following statistical relations
\begin{eqnarray}
\left\langle \zeta_{j}^{\alpha}(t)\right\rangle  & = & 0,\label{eq:fluc1}\\
\left\langle \zeta_{j}^{\alpha}(t)\zeta_{k}^{\gamma}(t')\right\rangle  & = & 2\eta\beta^{-1}\delta_{\alpha\gamma}\delta_{jk}\delta(t-t'),\label{eq:fluc2}
\end{eqnarray}
where $\alpha,\gamma\in\{x,z\}$ and $j,k\in\{1,2,...,N\}$.
\begin{figure}
\centering
\includegraphics[width=0.55\linewidth,angle=-90]{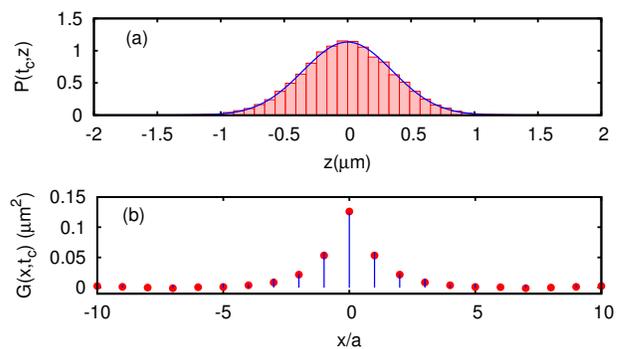}
\caption{{\small (a) Probability distribution for the transverse
displacement $z$ and (b) two-point correlation function $G(x,t)$ at the
critical point for a quench of $\tau_Q=41.7 \ \mu$s and $N=21$ ions. The
histogram and points correspond to the Langevin approach, averaging over
$2000$ stochastic trajectories, while solid lines to the Fokker-Planck
approach. See main text for further details regarding the used parameters.}}
\label{fig:8}
\end{figure}
We solve equations of motion, Eqs.~(\ref{eq:ion_motion})-(\ref{eq:ion_motion2}),
using the Langevin Impulse integration method with a timestep of $10$
ns~\cite{Skeel:02}. We consider a system of $N=21$ ions, with inter-ion
spacing in the linear configuration $a=10 \ \mu$m,  mass $m=172$ amu, which
corresponds to $\textrm{Yb}^+$ ions, temperature $T=5$ mK and friction
coefficient $\eta=1.5\times 10^{-21}$ kg s${}^{-1}$~\cite{Pyka:13}.
 The initial and final transverse frequency is set to
$\omega_i=2\pi\times477.5$ kHz and $\omega_f=2\pi\times 159$ kHz. Note that
$\omega_0=2\pi\times 143$ kHz and hence, the critical frequency is
$\omega_t^c=2\pi \times 293.4$ kHz. To ensure that the system is initially in
thermal equilibrium, the system is evolved under fixed trap parameters for
$100\ \mu$s before starting the quench protocol. 
The quench time $\tau_Q$ is varied from $10\ \mu$s to $400\ \mu$s. For each
value of $\tau_Q$, we perform $2000$ simulations in order to obtain accurate
estimations of statistical observables such as two-point correlation function
$G(x,t)$, correlation length $\xi(t)$, number of defects at the end of the
quench $\left<N_d \right>$, and probability distributions of the
transverse displacement $z(t)$, which due to translational symmetry does not depend 
on the ion position.

\begin{figure}
\centering
\includegraphics[width=0.51\linewidth,angle=-90]{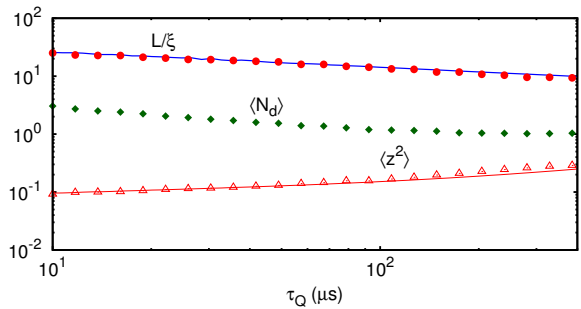}
\caption{{\small Scaling of correlation length $L/\xi$, mean square transverse displacement $\left< z^2\right>$ at the critical point, using Langevin approach, averaging over $2000$ stochastic trajectories (points) and Fokker-Planck formalism (solid lines), together with the average number of structural defects $\left<N_d \right>$ at the end of the quench, which saturates to $1$ for slow quenches. Note that $L/\xi\sim 10\left<N_d \right>$. A fit to a $\tau_Q^{\alpha}$ gives $\alpha=-0.31(1)$ and $-0.30(1)$ in the range of $\tau_Q\in\left[40,200\right]\ \mu$s for $L/\xi$ and $\left<N_d \right>$ using Langevin approach, respectively, while for Fokker-Planck results in $-0.29(1)$ for $L/\xi$.} }
\label{fig:9}
\end{figure}

\begin{figure}
\centering
\includegraphics[width=0.4\linewidth,angle=-90]{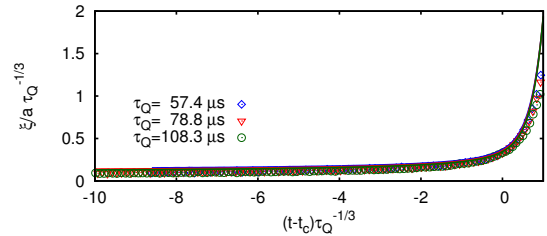}
\caption{{\small Collapse of $\xi$ into a single curve for $N=21$ ions evolving towards the critical point for three different quench times $\tau_Q$. Note that $t_c$ corresponds to the time at which the critical frequency is reached, i.e., $\omega_t(t_c)=\omega_t^c$. The points correspond to the Langevin approach, averaging over $2000$ stochastic trajectories, while solid lines to the Fokker-Planck approach.} }
\label{fig:10}
\end{figure}

Fig.~\ref{fig:8} shows an example of the results of the calculations for a
selected quench time $\tau_Q=41.7 \ \mu$s, namely the  probability distribution
of the transverse displacement and the two-point correlation function at the
critical point.  In Fig.~\ref{fig:9}, we plot the scaling of several
quantities as a function of the quench time $\tau_Q$. These quantities are
the correlation length $\xi$ and the averaged square displacement $\left<z^2
\right>$ at $\omega_t(t_c)=\omega_t^c$, and  the number of defects $\left<N_d
\right>$ at the end of the quench. We find that the scaling exponent is
$\sim1/3$, which is in agreement with the existing results in
Refs.~\cite{Chiara:10,Nigmatullin:16}, which predict this scaling by mapping
the problem to GL theory and using the KZ relation $\tau_Q^{-\nu/(1+z\nu)}$
with $\nu=1/2$ and $z=1$. Furthermore, following the theory developed in~\cite{Gor:16}, we can transform physical quantities to remove their dependence on the quench time $\tau_Q$, as explained and demonstrated in previous sections for GL theory.  This entails the collapse of the correlation length $\xi$ into a single curve for different $\tau_Q$ values when $\xi \tau_Q^{-1/3}$ is plotted  against the rescaled time $(t-t_c)\tau_Q^{-1/3}$ with $\omega_t(t_c)=\omega_t^c$, as shown in Fig.~\ref{fig:10}. Note that since  $\nu=1/2$ and $z=1$, the used transformation to obtain the collapse is equivalent as GL theory in underdamped regime (see Sec.~\ref{sec:kram}).
In addition to the results of the Langevin dynamics
simulations, Figs.~\ref{fig:8},~\ref{fig:9} and~\ref{fig:10} include the results of
the Fokker-Planck approach. The results show a good agreement with the
Fokker-Planck description of the problem even when non-linear terms are neglected, 
as we explain in the following.


\emph{Fokker-Planck approach.---} We now apply the Fokker-Planck approach to the problem of non-equilibrium
quenches from the linear to zigzag configuration. At the start of
the quench the system is in the symmetric linear phase. The equilibrium
configuration of the ions is given by $\textbf{r}_{j}^{(0)}=(x_{j}^{(0)},0)$,
where for convenience we take $x_{i}>x_{j}$ for $i>j$. Due to the
periodic boundary conditions, the equilibrium inter-particle distance
is constant i.e. $a=x_{j+1}^{(0)}-x_{j}^{(0)}$. The linearized equations
of motions for small displacements around the equilibrium configurations
$q_{j}=x_{j}-x_{j}^{(0)}$ and $z_{j}$ are obtained by Taylor expanding
the potential. In the second-order Taylor expansion the axial and
transverse motion decouple~\cite{Chiara:10}. The equations of motion for the transverse
displacements in this limit are
\begin{equation}
    m\ddot{z}_{j}+\eta\dot{z}_{j}+m\omega_{t}^{2}z_{j}-\frac{1}{2}\sum_{i\neq j}\mathcal{K}_{i,j}\left(z_{j}-z_{i}\right)=\zeta_{j}^z(t)\label{eq:transversmotion}
\end{equation}
where $\mathcal{K}_{i,j}\equiv\left.-\partial^{2}V/\partial x_{j}\partial x_{i}\right|_{x_{j}^{(0)}}$
is given by
\begin{equation}
    \mathcal{K}_{i,j}=\frac{2Q^{2}}{\left|x_{i}^{(0)}-x_{j}^{(0)}\right|^{3}}.\label{eq:kappa}
\end{equation}

Eq.~(\ref{eq:transversmotion}) describes the motion of coupled
oscillators, which can be decoupled by rewriting it
in terms of the normal modes. The relation between the transverse coordinate
vector $\vec{z}$  and the normal mode vector $\vec{\Psi}$
can be written as
\begin{align}
    &z_{j}=\frac{1}{\sqrt{N}}\Psi_0^{+}+\nonumber \\
&+\sqrt{\frac{2}{N}}\sum_{n=1}^{(N-1)/2}\left(\Psi_{n}^{+}\cos(k_n j a)+\Psi_{n}^{-}\sin(k_n j a)\right),\label{eq:normal_mode}
\end{align}
where $k_n=2\pi n/Na$ and the sign $+$ ($-$) indicates the parity under
$k_n\rightarrow -k_n$. Substituting Eq.~(\ref{eq:normal_mode})
in~(\ref{eq:transversmotion}) gives
\begin{equation}
    m\ddot{\Psi}^{\pm}_{n}+\eta\dot{\Psi}^{\pm}_{n}+m\mbox{\ensuremath{\omega}}_{n}^{2}(t)\Psi_{n}^{\pm}=\zeta_{n}^{\pm}(t),\label{eq:modes}
\end{equation}
where $\omega_n(t)$ defines the frequency of the normal modes,
\begin{equation}
    \omega_{n}^{2}(t)=\omega_{t}^{2}(t)-2\left(\frac{2Q^{2}}{ma^{3}}\right)\sum_{j=1}^{N}\frac{1}{j^{3}}\sin^{2}\left(\frac{k_n j a}{2}\right),\label{eq:spectrum}
\end{equation}
and $\zeta_n^{\pm}(t)$  represents the stochastic force in the normal mode space, which again fulfills $\left<\zeta_n^{p}(t)\zeta_m^{q}(t') \right>=2\eta/\beta\delta_{pq}\delta_{nm}\delta(t-t')$.
The Fokker-Planck equations corresponding to the Langevin Eq.~(\ref{eq:modes}) are
\begin{align}
&    \frac{\partial P_{n}(t,\Psi_{n},\dot{\Psi}_{n})}{\partial t} =  \left[-\frac{\partial}{\partial\Psi_{n}}\dot{\Psi}_{n}+\frac{\eta}{2\beta m^2}\frac{\partial^{2}}{\partial\dot{\Psi}_{n}^{2}}+\right. \nonumber\\
&+\left.\frac{\partial}{\partial\dot{\Psi}_{n}}\left(\frac{\eta}{m}\dot{\Psi}_{n}+\omega_{n}^{2}(t)\Psi_{n}\right)\right]P_{n}(t,\Psi_{n},\dot{\Psi}_{n}).\label{eq:modeFP}
\end{align}

Therefore, we have reduced the problem to the solution of $N$ deterministic
Fokker-Planck equations that determine the mode population probabilities
at a chosen time $t$. Using Eq.~(\ref{eq:modeFP}) and the expressions for
$P_{n}$, following the same procedure as in Sec.~\ref{sec:kram}, allows the
determination of any statistical observable, as well for example, the
probability distributions for $z_j$ at time $t$. Note that, from Eq.~(\ref{eq:normal_mode}) and since $\Psi_n^{\pm}$ are statistically independent and Gaussian distributed, $P(t,z_j)$ adopts also a Gaussian form and independent of $j$,
\begin{align}
P(t,z)=\frac{1}{\sqrt{2\pi\sigma^2(t)}}e^{-z^2/(2\sigma^2(t))},
\end{align}
 with a time-dependent variance 
\begin{align}
\sigma^2(t)=\frac{1}{N} \left( \frac{1}{2F_0^2(t)}+\sum_{n=1}^{(N-1)/2} \frac{1}{F_n^2(t)}\right),
\end{align}
where $F_n(t)$ is obtained from Eq.~(\ref{eq:modeFP}) in the same way as explained in Sec.~\ref{sec:kram}.  In
particular, we calculate the non-equilibrium correlation length $\xi$ and the
mean square transverse displacement $\left<z^2\right>$ for the same set of
parameters as was used previously in the Langevin approach. A comparison
between the two approaches is shown in Fig.~\ref{fig:8} and~\ref{fig:9}. We
emphasize that the Fokker-Planck results which have been obtained under a 
simplified description of the realistic model, where non-linear terms and fluctuations in the
longitudinal coordinates have been neglected, still reproduce essential
features of the considered non-equilibrium scenario in a quantitative way.

\section{Conclusions}
\label{sec:conc}

We have studied the emergence of universal scaling laws in non-equilibrium second-order phase transitions using Fokker-Planck formalism. We verify that the developed approach reproduces Kibble-Zurek scaling laws in one dimensional Ginzburg-Landau model in overdamped and underdamped dynamical regimes. Additionally, we use this approach to obtain the universal finite-size scaling functions and demonstrate the universality of the dynamics.

There are several advantages of the developed method. It allows us to determine universal scaling laws in a efficient way in comparison to Langevin approach, where ensemble averages must be computed numerically. It provides analytic results that are easily amenable to further analysis. Moreover, it has an extended range of applicability - it can be used to derive insights into the non-equilibrium symmetry breaking phase transitions in overdamped, underdamped and intermediate dynamical regimes; finite as well as infinite systems and systems that are not described directly by the Ginzburg-Landau model.
We have illustrated the power of the developed framework by analyzing the non-equilibrium linear to zigzag structural phase transition of an ion chain with periodic boundary conditions. We find an excellent agreement between the results obtained using the Fokker-Planck approach and the non-linear Langevin dynamics simulations.

One challenge for future theoretical work is to include the coupling between the normal modes of the system during quench protocol and find a way to predict the resulting corrections to the scaling laws. Another interesting direction of research would be apply this framework to investigation of scaling laws of other important quantities in stochastic thermodynamics such as entropy production and work done.

\acknowledgments{
 This work is  supported by an Alexander von Humboldt Professorship, by DFG through grant ME 3648/1-1 and the EU STREP project EQUAM. This work was performed on the computational resource bwUniCluster funded by the Ministry of Science, Research and the Arts Baden-W\"urttemberg and the Universities of the State of Baden-W\"urttemberg, Germany, within the framework program bwHPC. R. P. thanks P. Fern\'andez-Acebal and A. Smirne for useful discussions.}

\end{document}